\documentclass[11pt]{article}
\usepackage{amsfonts}
\usepackage{amsthm}
\usepackage{amscd}
\usepackage{amsmath}

 \usepackage{verbatim}
\usepackage{enumerate}
 
\usepackage{amssymb}

\usepackage{color}
\usepackage{graphicx}
\usepackage{graphics}
 \usepackage{url}

\newcommand\bx {\mathbf x}

\newcommand\bA {\mathbf A}
\newcommand\bB {\mathbf B}

\newcommand\bthe {\mbox{\boldmath $\theta$}}

\def\real{\mathbb{R}}

\def\square{\ifmmode\sqr\else{$\sqr$}\fi}
\def\sqr{\vcenter{
         \hrule height.1mm
         \hbox{\vrule width.1mm height2.2mm\kern2.18mm
\vrule width.1mm}
         \hrule height.1mm}}

\begin{document}

\title{On the instability of two  entropic dynamical models\thanks{\scriptsize   This research was partially supported by Grants 20020120200244BA from the Universidad de Buenos Aires, \textsc{pip} 11220110100742  from \textsc{conicet} and \textsc{pict}-2012-1641 from \textsc{anpcyt}, Argentina.}}
\author{Guillermo Henry and   Daniela Rodriguez\\
{\small  \sl Facultad de Ciencias Exactas y Naturales, Universidad de Buenos Aires and
CONICET, Argentina.}}
\date{}
\maketitle

\begin{abstract}
In this paper we study two entropic dynamical models from the viewpoint of information geometry. We study  the geometry structures of the associated statistical manifolds. In order to analyse the character of the instability of the systems, we  obtain  their geodesics  and compute their Jacobi vector fields. The results of this work improve and extend a recent advance in this topics studied in \cite{peng}.
\end{abstract}

\section{Introduction}\label{intro}

%p.8:. cuál es la generalidad de los supuestos realizados entre la ecuación (15) y la ecuación (16)

The evolution of some  systems could be predicted with certitude, however in some cases, by the complexity of the system, lack of information, etc, the  predictions of final states can be done at the best only by assigning probabilities. Examples of these system could be found in biology, ecology, chemistry, physics, and  economics.
Some authors believe that  quantum mechanics might be derived by the laws of probability inference, as well as happens with  thermodynamic (see for instance \cite{cafaro2} and \cite{caticha}).  
 Entropic Dynamics (see \cite{caticha2}) provided a tool that could be useful in the study of the dynamics of certain complex systems. Roughly, given a system, the Entropic Dynamic make use of maximum relative entropy principle in order to determine  a statistical manifold that model it. This statistical manifold represent the total macro-states of the system (i.e.,  probability distributions). To obtain this manifold, firstly  we have to determine the micro-states and the constraint of the system. For instance, if we  want to study the dynamics of $k$ particles in a $l$-dimensional Euclidean space, the micro-states could be the ${lk}$-random  variables $\bx=(x_1, \dots, x_{k})\in \real^{lk}$  with $x_i=x_i^1,\dots,x^l_i$ and distributions $p_{i}^{j}$ that represent the position of the particles. The constraints could be the expected values or the variances of $p_i^j$, or some extra knowledge, for instance, if these distributions are correlated or not. These constraints are the only testable information that we can get from the system. In order to get the family of distributions that better fit to the system we maximize the relative entropy functional (see \cite{caticha}) given a prior probability density (the uniform distribution). In the case that the constrains are the expected valued $u_i^j$ and the variance $v^j_i$ of $p_i^j$ and assuming  that $x^j_i$ are independent distributed random variables, then we will get a statistical manifold $S$ of dimension $2lk$ parametrized  by a function $\phi$ over some open set of $\real^{2lk}$ 

$$\Big((u^1_1,v^1_1),\dots,(u_k^l,v_k^l)\Big)\longrightarrow \phi\Big((u^1_1,v^1_1),\dots,(u_k^l,v_k^l)\Big)=(p_1^1,\dots,p_1^l,\dots,p_k^1,\dots, p_k^l)\in S.$$

We are going to consider the geometry of the manifold $S$ induced by the Fisher information metric $g$ (see section \ref{kmanifold} for the definition).  The evolution of the system can be seen as a continuous path in $S$. The entropic dynamics principle claims that the system evolves followings the geodesics of the Riemannian manifold $(S,g)$. Therefore,  the curvature of $(S,g)$ encoded some information on the dynamic of the system. So, the task is  to study the geometry of $(S,g)$ from the Information Geometry  viewpoint (see \cite{amari2} and \cite{arwini}) in order to understand the features of the system under consideration.    
There are several references related with the study of entropic dynamical models from the viewpoint of information geometry, see for instance  \cite{cafaro}, \cite{cafaro2}, \cite{peng2} among others.

Nevertheless, there does not exist a general standard procedure to set up the appropriated constraints for a given system.  Most of the time this must be done by intuition or by some experimental data. So, it seen  
important to understand the geometry of some statistical models. In the present article we study some statistical manifolds that appear in several fields, such as  physics, biology, social sciences, economics, see for instance  \cite{guerriero}, \cite{rosin}, \cite{sagias}, \cite{wei}, \cite{yuji}, \cite{eltoft}, and \cite{reed}, among others.

The aim of present article is to extend and study two entropic dynamical models introduced by Peng, Sun, Sun, and Yi  in \cite{peng}.

  In \cite{peng}, the authors studied the character of the instability of   two entropic dynamical models:
  \begin{itemize}\item${{M}}_1$: with a statistical manifold induced by a family  of a  joint Gamma and Exponential distributions \item ${{M}}_2$: with a statistical manifold induced by a family of a joint Gamma and Gaussian distributions.\end{itemize} 
  From the study of the geometry of both models, they found out that $M_1$ have first order linear divergent instability and $M_2$ have  exponential instability.
  
The first model that we consider is given by the statistical manifold induced by the   $k-$joint one parametric exponential family (it model a system of uncorrelated $k$ particles). Second, we study a system of two correlated particles modelled by the statistical manifold of the multivariate Gaussian probability family.
Finally, we discuss how these models can be combined in order to generalize the obtained results to a large class of models. 
 
The  paper is organized as follows. In Section 2, we introduce  a $k$-dimensional statistical manifold induced by  densities of a one parameter exponential family and  we study its geometrical structure. We analyse  the character of the stability of this model when $k=4$. In section 3, we  study the geometric structure and the stability of a Gaussian statistical manifolds with correlations.  Conclusions and some extensions are presented in Section 4.

\section{Geometric structure and stability of $k-$dimensional statistical manifold}\label{kmanifold}
We refer the reader to \cite{amari2} and \cite{murray}  for definitions and standard results  concerning to the geometry of statistical manifolds. 

We consider a system of $k$ particles in a one dimensional space named $\bx=(x_1,\dots,x_k)$.  We assume that all information relevant to the dynamical model comes from the probability distribution which in this case is the joint distribution of $k$ independent one parameter exponential family. More precisely, we consider the following joint density function
$$p(\bx,\bthe)=  h(\bx) \exp\left(\sum_{s=1}^k (\eta_s(\theta_s)T_s(x_s)-\gamma_s(\theta_s))\right)$$
with $\bthe=(\theta_1,\dots,\theta_k)$, $\bx=(x_1,\dots,x_k)$, $ T_s$ is a continuous function  and  $\eta_s$ and $\gamma_s$ are twice-differentiable functions   for $s=1,\dots, k $.
Therefore, we can define the associated statistical manifold as follows
$${M}_k:=\left\{ p(\bx,\bthe)=  h(\bx) \exp\left(\sum_{s=1}^k (\eta_s(\theta_s)T_s(x_s)-\gamma_s(\theta_s))\right) \;\; \theta_s\in \real \mbox{ for } s=1,\dots, k  \;\right\}.$$
We are going to consider $M_k$ endowed with the Fisher-information matrix. This metric is proportional to the amount of information that the distribution function contains about the parameter. Recall that the local expression of the  Fisher-information metric with respect to the coordinate system  $\bthe$  is:
$$g_{ij}(\bthe)  = E\left(\partial_i l(\bthe)\partial_j l(\bthe)\right)$$
where $\partial_i l (\theta)=\frac{\partial}{\partial \theta_i}\log p(\bx,\bthe)$.  It is easy to see that the Fisher-information metric on ${M}_k$ can be computed as 
$$
g_{ij}(\bthe)=E\left((\eta^{\prime}_i(\theta_i)T_i(x_i)-\gamma_i^{\prime}(\theta_i))(\eta^{\prime}_j(\theta_j)T_j(x_j)-\gamma_j^{\prime}(\theta_j))\right).
$$
Since  the variables $x_s$  ($s=1,\dots, k$) have  density function belonging to one parameter exponential family,  the expected value and the variance of $T_s$ can be computed easily in terms of  $\eta_s$ and $\gamma_s$. Indeed, 
$$
E( T_s)=\frac{\gamma_s^{\prime}(\theta_s)}{\eta_s^{\prime}(\theta_s)}\quad\quad Var (T_s)=\frac{\gamma_s^{\prime\prime}(\theta_s)\eta_s^{\prime}(\theta_s)-\gamma_s^{\prime}(\theta_s)\eta_s^{\prime\prime}(\theta_s)}{(\eta_s^{\prime}(\theta_s))^3}.
$$
From the independence of the variables $x_s$ we have 
$$
g_{ij}(\bthe)=\delta_{ij}(\eta^{\prime}_i(\theta_i))^2\;Var(T_i)=\delta_{ij}\frac{\gamma_i^{\prime\prime}(\theta_i)\eta_i^{\prime}(\theta_i)-\gamma_i^{\prime}(\theta_i)\eta_i^{\prime\prime}(\theta_i)}{\eta_i^{\prime}(\theta_i)},
$$
where $\delta_{ij}$ is the Kronecker's delta.  Note that we have assumed uncoupled constraints between the micro-variables.  This assumptions leads to a metric tensor with trivial off diagonal elements.

The inverse matrix of $g$ is
$$g^{-1}=[g^{ij}]=diag\left(\frac{\eta_1^{\prime}(\theta_1)}{\gamma_1^{\prime\prime}(\theta_1)\eta_1^{\prime}(\theta_1)-\gamma_1^{\prime}(\theta_1)\eta_1^{\prime\prime}(\theta_1)},\dots,\frac{\eta_k^{\prime}(\theta_k)}{\gamma_k^{\prime\prime}(\theta_k)\eta_k^{\prime}(\theta_k)-\gamma_k^{\prime}(\theta_k)\eta_k^{\prime\prime}(\theta_k)}\right).$$
The length element is given by 
$$
ds^2= g_{ij}d\theta_i\theta_j=\sum_i\frac{\gamma_i^{\prime\prime}(\theta_i)\eta_i^{\prime}(\theta_i)-\gamma_i^{\prime}(\theta_i)\eta_i^{\prime\prime}(\theta_i)}{\eta_i^{\prime}(\theta_i)} \;d\theta_i^2,
$$
and the volume  element  is
\begin{equation}\label{eqvol}
d{V}_{g}= \sqrt{g}\;d\theta_1\wedge\dots \wedge d\theta_k=\left(\prod_i\frac{\gamma_i^{\prime\prime}(\theta_i)\eta_i^{\prime}(\theta_i)-\gamma_i^{\prime}(\theta_i)\eta_i^{\prime\prime}(\theta_i)}{\eta_i^{\prime}(\theta_i)}\right)^{1/2} \;d\theta_1\wedge\dots \wedge d\theta_k.
\end{equation}
where $ \sqrt{g}=\sqrt{\det(g_{ij})}$. 

Recall that the Christoffel symbols $\Gamma_{ij}^l$ is defined by
$\Gamma_{ij}^l=\Gamma_{ijs}g^{sl}$ ($i,j,l,s=1,2,\dots,k$)
where $$
\Gamma_{ijs}=\frac 12(\partial_ig_{js}+\partial_jg_{si}-\partial_sg_{ij}) , \quad\quad i,j,s=1,\dots,k.
$$
For this model the  Christoffel symbols that are not zero are:

\begin{eqnarray}\label{chris}
\Gamma_{ii}^i&=&\frac{\gamma_i^{\prime\prime\prime}(\theta_i)(\eta_i^{\prime}(\theta_i))^2-\gamma_i^{\prime}(\theta_i)\eta_i^{\prime\prime\prime}(\theta_i)\eta_i^{\prime}(\theta_i)-\eta_i^{\prime\prime}(\theta_i)(\gamma_i^{\prime\prime}(\theta_i)\eta_i^{\prime}(\theta_i)-\gamma_i^{\prime}(\theta_i)\eta_i^{\prime\prime}(\theta_i))}{2\eta_i^{\prime}(\theta_i)(\gamma_i^{\prime\prime}(\theta_i)\eta_i^{\prime}(\theta_i)-\gamma_i^{\prime}(\theta_i)\eta_i^{\prime\prime}(\theta_i))}\nonumber\\
&=&\frac12 \left(\frac{\gamma_i^{\prime\prime\prime}(\theta_i)\eta_i^{\prime}(\theta_i)-A_i^{\prime}(\theta_i)\eta_i^{\prime\prime\prime}(\theta_i)}{A_i^{\prime\prime}(\theta_i)\eta_i^{\prime}(\theta_i)-\gamma_i^{\prime}(\theta_i)\eta_i^{\prime\prime}(\theta_i)}-\frac{\eta_i^{\prime\prime}(\theta_i)}{\eta_i^{\prime}(\theta_i)}\right).
\end{eqnarray}

The Ricci curvature $R_{is}$ is defined by $R_{is}=R_{ijsl} g^{jl} \; i,j,s,l=1,\dots,k$ where
$$
R_{ijsl}=(\partial_j\Gamma_{is}^u-\partial_i\Gamma_{js}^u)g_{ul}+(\Gamma_{jtl}\Gamma_{is}^t-\Gamma_{itl}\Gamma_{js}^t).
$$
Therefore, it is easy to see that the curvature tensor components are all zero and  the scalar curvature $S_{g}=0$.

Recall that the geodesic equations  are given by the following  non linear system of second order ordinary differential equations:

\begin{equation} \label{eqgeo}
\frac{\partial^2\theta_l}{\partial \tau^2}+\Gamma_{ij}^l\frac{\partial \theta_i}{\partial \tau}\frac{\partial \theta_j}{\partial \tau}=0\quad \quad \mbox{ for } i,j,l=1\dots,k.
\end{equation}
From (\ref{chris}) we obtain that the geodesics are determined by the following  $k$ differential equations:
\begin{eqnarray}
\frac{\partial^2\theta_i}{\partial^2 \tau}+\Gamma_{ii}^i\left(\frac{\partial \theta_i}{\partial \tau}\right)^2=0\quad\quad\mbox{for} \quad\quad i=1,\dots,k.\label{eqkm}
\end{eqnarray}

\noindent\bf Remark \ref{kmanifold}.1: \rm The entropic dynamical model ${{M_1}}$ studied in  \cite{peng} (Section 3) is a particular case of  the model introduced in this section. More precisely, taken $k=2$ ,  $\eta_1(\theta_1)=\frac{-\rho}{\theta_1}$, $\eta_2(\theta_2)=\frac{-1}{\theta_2}$, $\gamma_1(\theta_1)=-\rho\ln(\theta_1)$, $\gamma_2(\theta_2)=-\ln(\theta_2)$ and $h(\bx)=\frac{1}{\Gamma(\rho)}x_1^{\rho-1}$ we get ${{M_1}}$. Therefore, the results given in this section extend those ones obtained in   \cite{peng} .

\subsection{Instability}\label{sec21}

In this section we consider a system of $4$ particles in a one dimensional space. We assume that the particles $\bx=(x_1,\dots,x_4)$ (the micro-states) do not interact between them and are distributed according  to  Poisson, Pareto, Laplace, and  Weibull distributions, respectively. More precisely, the joint probability density function  is 
\begin{equation}\label{modelo1}
p(\bx,\bthe)= \frac{\theta_1^{x_1}e^{-\theta_1}}{x_1!}\;\theta_2a^{\theta_2}x_2^{-(\theta_2+1)}\; \frac{1}{2\theta_3}e^{-\frac{|x_3|}{\theta_3}}  \;\frac{bx_4^{b-1}}{\theta_4^b}e^{-\frac{x^b_4}{\theta_4^b}}
\end{equation}
with  $\theta_i>0$ for $i=1,\dots,4$. This is a particular case of the entropic dynamic model introduced in  the previous section. Indeed, taking $k=4$ and 
$$
\begin{array}{ll}
\eta_1(\theta_1)=\ln(\theta_1)& \gamma_1(\theta_1)=\theta_1\\
\eta_2(\theta_2)=-(\theta_2+1)& \gamma_2(\theta_2)=-\ln(\theta_2)-\theta_2\ln(a)\\
\eta_3(\theta_3)=-\frac{1}{\theta_3}&\gamma_3(\theta_3)=\ln(2\theta_3)\\
\eta_4(\theta_4)=-\frac{1}{\theta_4^b}& \gamma_4(\theta_4)=b\ln(\theta_4)-\ln(b)\\
\end{array}
$$
($a$ and $b$ are fixed and known) $M_4$ modelled the system described above. From (\ref{eqkm}) we get that the geodesic equations are

\begin{eqnarray}\label{geodesicmodel1}
\frac{\partial^2\theta_1}{\partial^2 \tau}&=&\frac{1}{2\theta_1}\left(\frac{\partial \theta_1}{\partial \tau}\right)^2\\
\frac{\partial^2\theta_i}{\partial^2 \tau}&=&\frac{1}{\theta_i}\left(\frac{\partial \theta_i}{\partial \tau}\right)^2\qquad\mbox{ for } i=2,3,4,
\end{eqnarray}
whose solution is
\begin{equation}\label{geom1}
\theta_1(\tau)=A_1\left(t+B_1\right)^2 \quad\quad \theta_i(\tau)=A_ie^{B_i\tau} \quad\mbox{for}\quad i=2,3,4\quad A_i\in\real-\{0\}, B_i\in\real.
\end{equation}
Let $\bA=(A_1,\dots,A_4)$ ($A_i\neq 0$) and $\bB=(B_1,\dots,B_4)$. We denote with $\alpha_{\bA,\bB}$ the geodesic obtained from replace $\bA$ and $\bB$ in (\ref{geom1}). The arc-length of $\alpha_{\bA,\bB}$ between  $\alpha_{\bA,\bB}(0)$ and $\alpha_{\bA,\bB}(\tau)$ is 
\begin{eqnarray*}
\ell^{\tau}(\alpha_{\bA,\bB})&=&\int_0^{\tau}\left(\sum_i g_{ii}\left(\frac{\partial \theta_i}{\partial \tau}\right)^2\right)^{1/2}\;ds=\int_0^{\tau}\left(4A_1+B_2^2+B_3^2+b^2B_4^2\right)^{1/2}\;ds\\
&=&\left(4A_1+B_2^2+B_3^2+b^2B_4^2\right)^{1/2}\tau.
\end{eqnarray*}
Note that  the geodesic length is independent of  $A_2,A_3,A_4$  and $B_1$. The difference of the length of two geodesics with close initial condition diverges. For instance,  
$$D(\tau)=|\ell^{\tau}(\alpha_{(A_1+\delta,\dots,A_4),\bB})-\ell^{\tau}(\alpha_{\bA,\bB})|\longrightarrow_{\tau\to\infty} +\infty.$$

From (\ref{eqvol}), we have that the volume element is
$$
d{ V}_{g}=\left(\frac{1}{\theta_1}\right)^{1/2}\frac{1}{b\theta_2\theta_3\theta_4}d\theta_1\wedge d\theta_2\wedge d\theta_3\wedge d\theta_4.
$$
Thus the volume of an extended region of ${M}_4$ is
$$ 
\Delta V_{{M}_4}(\tau)=\int_{\theta_1(0)}^{\theta_1(\tau)}\int_{\theta_2(0)}^{\theta_2(\tau)}\int_{\theta_3(0)}^{\theta_3(\tau)}\int_{\theta_4(0)}^{\theta_4(\tau)}d{ V}_{g}=2\sqrt{A_1}B_2B_3B_4(\tau+B_1)\tau^3$$
and  the average volume is
$$
\frac1{\tau}\int \Delta V_{{ M}_4}(\tau)= 2\sqrt{A_1}B_2B_3B_4(\frac{\tau}{5}+\frac{B_1}{4})\tau^3.
$$
This quantity  encodes relevant information about the stability of neighbouring volume region. The asymptotic behaviour of the average volume  has diffusive expansion that  increase as a polynomial function.  

Finally, we study the temporal behaviour of the Jacobi field equation which is a  natural tool to analyse  dynamical chaos (analysing the geodesic spread). First, we recall that the Jacobi field   equation (see \cite{docarmo}) is 
\begin{eqnarray}\label{jlc}
\frac{D^2J_i}{D\tau^2}=R^i_{kml}\frac{\partial \theta^k}{\partial \tau}J_m\frac{\partial \theta^l}{\partial \tau}
\end{eqnarray}
with $J=(J_1,J_2,J_3,J_4),$ $R^i_{kml}=\partial_m\Gamma^i_{kl}-\partial_k\Gamma_{ml}^i+\Gamma^j_{kl}\Gamma^i_{mk}- \Gamma_{ml}^j\Gamma_{kj}^i$ %, $\delta\theta^i=\delta_{\alpha}\theta^i:=\left(\frac{\partial  \theta^i(\tau;\alpha)}{\partial \alpha}\right)\delta \alpha$  where $\theta^i$ are the solutions of (\ref{eqgeo})
and the covariant derivative is defined as follows
$$
\frac{D^2J_i}{D\tau^2}=\frac{\partial^2J_i}{\partial\tau^2}+2\Gamma^i_{jk}\frac{\partial J_j}{\partial\tau}\frac{\partial\theta^k}{\partial\tau}+
\Gamma^i_{jk}J_j\frac{\partial^2\theta^k}{\partial^2\tau}+\partial_h\Gamma^i_{jk}\frac{\partial\theta^h}{\partial\tau}\frac{\partial\theta^k}{\partial\tau} J_j+\Gamma^i_{jk}\Gamma^j_{ts}\frac{\partial\theta^s}{\partial\tau}\frac{\partial\theta^k}{\partial\tau}J_t.
$$

In our case, the fact that the entropy dynamical model is uncoupled implies that $R^i_{kml}=0$ for all $i,k,m,l$.  Therefore  the relative geodesic spread characterized by the Jacobi field equation is given by the following set of second order differential equations:
\begin{eqnarray*}
(\tau+B_1)^2\frac{\partial^2J_1}{\partial\tau^2}-2(\tau+B_1)\frac{\partial J_1}{\partial\tau}+2J_1&=&0\\
\frac{\partial^2J_i}{\partial\tau^2}-2B_i\frac{\partial J_i}{\partial\tau}+B_i^2J_i&=&0\quad\mbox{ for }\quad i=2,3,4.
\end{eqnarray*}
Hence the coordinates of the Jacobi field are given by :
\begin{eqnarray*}
J_1(\tau)&=&a_{1,1}(\tau+B_1)\\
J_i(\tau)&=&(a_{1,i}+a_{2,i}\tau) e^{B_i\tau}\quad\mbox{ for }\quad i=2,3,4,
\end{eqnarray*}
where $a_{i,j}$ are integration constants. From this we can compute the square norm of the Jacobi field ($\|J\|^2=\sqrt{g_{ij}J_iJ_j}$). We have that
$$
\|J\|^2=\frac{a_{1,1}^2}{A_1}+\left(\frac{a_{1,2}+a_{2,2}\tau}{A_2}\right)^2+\left(\frac{a_{1,3}+a_{2,3}\tau}{A_3}\right)^2+\left(b\,\frac{a_{1,4}+a_{2,4}\tau}{A_4}\right)^2.
$$
This shows that the  Jacobi vector field intensity diverges polynomially.

\section{Geometric structure and stability of Gaussian statistical manifold with correlations}\label{gaussiana}
In the previous sections, we analysed the geometry and the instability of a model of $k$ particles with no interaction between them. In this section, we will consider two particles that interact between them with a certain correlation. More precisely, we will consider a Gaussian statistical manifold in  presence of correlations. Recall that  the density function of two random variables with joint Gaussian distribution is given by
\begin{equation}\label{modelo2}
p(\bx,\bthe)=\frac{1}{2\pi\sigma^2\sqrt{1-r^2}}\exp\left\{-\frac{1}{2\sigma^2(1-r^2)}\left[(x-\mu_x)^2-2r(x-\mu_x)(y-\mu_y)+(y-\mu_y)^2\right]\right\},
\end{equation}
where $\bthe=(\mu_x,\mu_y,\sigma)$,  $x,y\in\real$ and   $|r|< 1$ is a known parameter. Let us denote with ${M}_{G}$ the statistical manifolds associated to $p$ given by 
$${M}_{G}=\{ p(\bx,\bthe): \mbox{ with } \mu_x,\mu_y\in\real,\quad\sigma>0\}.$$
Note that the system is really different than the one we have analysed in the previous section. The coupled constraints would lead to a metric tensor with non-trivial off-diagonal elements given by the covariance terms.
We compute the matrix of the Fisher-information metric $[g_{ij}]_{{M}_{G}}$ and its inverse $[g^{ij}]_{{M}_{G}}$
$$
\begin{array}{cccc}
[g_{ij}]=
\frac1{\sigma^2}\left(\begin{array}{ccc}
\frac{1}{1-r^2}&\frac{-r}{1-r^2}&0\\
\frac{-r}{1-r^2}&\frac{1}{1-r^2}&0\\
0&0&4
\end{array}\right)
&&&[g^{ij}]=
{\sigma^2}\left(\begin{array}{ccc}
{1}&r&0\\
r&1&0\\
0&0&\frac14
\end{array}\right).
\end{array}
$$
Then the non zero coefficients $\Gamma_{ijk}$ of the Levi-Civita connection are:
$$
\begin{array}{lllll}
\Gamma_{113}=\Gamma_{223}=\frac{1}{(1-r^2)\sigma^3}&&\Gamma_{131}=\Gamma_{232}=\Gamma_{311}=\Gamma_{322}=\frac{-1}{(1-r^2)\sigma^3}&&\Gamma_{333}=\frac{-4}{\sigma^3}\\
\Gamma_{213}=\Gamma_{123}=\frac{-r}{(1-r^2)\sigma^3}&&\Gamma_{132}=\Gamma_{231}=\Gamma_{312}=\Gamma_{321}=\frac{r}{(1-r^2)\sigma^3}&&.\\
\end{array}
$$
Therefore,  the non zero Christoffel symbols are:
$$
\begin{array}{lll}
\Gamma_{11}^3=\Gamma_{22}^3=\frac{1}{4(1-r^2)\sigma}&&\Gamma_{33}^3=\Gamma_{13}^1=\Gamma_{23}^2=\Gamma_{31}^1=\Gamma_{32}^2=\frac{-1}{\sigma}\\
\Gamma_{21}^3=\Gamma_{12}^3=\frac{-r}{4(1-r^2)\sigma}&& %\Gamma_{33}^3=\frac{-1}{\sigma}.
\end{array}.
$$
The  non zero components of the curvature tensor  are: 
$$
\begin{array}{lll}
R_{1212}=R_{2121}=\frac{-1}{4(1-r^2)\sigma^4}&&R_{1221}=R_{2112}=\frac{1}{4(1-r^2)\sigma^4}\\
R_{1323}=R_{2313}=R_{3132}=\frac{r}{(1-r^2)\sigma^4}&&R_{1313}=R_{2323}=R_{3131}=\frac{-1}{(1-r^2)\sigma^4}.\\
\end{array}
$$
The components of the Ricci curvature are:
$$R_{11}=R_{22}=\frac{-1}{2(1-r^2)\sigma^2},\quad R_{12}=R_{21}=\frac{r}{2(1-r^2)\sigma^2}\quad\mbox{ and } \quad R_{33}=\frac{-2}{\sigma^2}.$$
 From this we  conclude that ${M}_{G}$ is a  manifold of constant  negative scalar curvature. More precisely, $
S_{{M}_{G}}=-\frac32
$. The sign of the scalar curvature is an expression of  chaos.  Negative  scalar curvature is a sufficient condition for the presence  of local instability. 

The geodesic equations for this model are:

\begin{equation}\label{eqg}
\left\{
\begin{array}{l}
\frac{\partial^2\mu_x}{\partial^2\tau}=\frac{2}{\sigma}\frac{\partial\mu_x}{\partial\tau}\frac{\partial\sigma}{\partial\tau}\\
\frac{\partial^2\mu_y}{\partial^2\tau}=\frac{2}{\sigma}\frac{\partial\mu_y}{\partial\tau}\frac{\partial\sigma}{\partial\tau}\\
\frac{\partial^2\sigma}{\partial^2\tau}=\frac{1}{\sigma}\left(\frac{\partial\sigma}{\partial\tau}\right)^2-\frac{1}{4(1-r^2)\sigma}\left(\left(\frac{\partial\mu_x}{\partial\tau}\right)^2+\left(\frac{\partial\mu_y}{\partial\tau}\right)^2\right)+\frac{2r}{4(1-r^2)\sigma}\frac{\partial\mu_x}{\partial\tau}\frac{\partial\mu_y}{\partial\tau}.
\end{array}
\right.
\end{equation}
A set of solutions of the system (\ref{eqg}) is given by
\begin{equation}\label{sol}
\mu_x(\tau)=C_x\frac{1}{1+e^{2C\tau}},\quad\quad\mu_y(\tau)=C_y\frac{1}{1+e^{2C\tau}}\quad\mbox{ and }\quad\sigma(\tau)=\frac{e^{{C}\tau}}{1+e^{2C\tau}}
\end{equation}
where $C=\frac14\sqrt{\frac{C_x^2+C_x^2-2rC_xC_y}{1-r^2}}$.

Let $\alpha_{C_x,C_y,r}$ be the geodesic obtained from replace $C_x$, $C_y$ and $r$ in (\ref{sol}). The arc-length of $\alpha_{C_x,C_y,r}$ between  $\alpha_{C_x,C_y,r}(0)$ and $\alpha_{C_x,C_y,r}(\tau)$ is given by 

\begin{eqnarray*}
\ell^\tau(\alpha_{C_x,C_y,r})&=&\int_0^\tau\frac{1}{\sigma}\left(\frac{1}{1-r^2}\left(\left(\frac{\partial\mu_x}{\partial\tau}\right)^2+ 	 \left(\frac{\partial\mu_y}{\partial\tau}\right)^2\right)-\frac{2r}{1-r^2}\left(\frac{\partial\mu_x}{\partial\tau}\right)\;\left(\frac{\partial\mu_y}{\partial\tau}\right)+4 \;\left(\frac{\partial\sigma}{\partial\tau}\right)^2\right)^{1/2}\;ds\\
&=& \int_0^\tau\left(\frac{1}{1-r^2}((C_x)^2+(C_y)^2)\sigma^2-\frac{2r}{1-r^2}C_x\;C_y\sigma^2+\frac{4}{\sigma^2} \;\left(\frac{\partial\sigma}{\partial\tau}\right)^2\right)^{1/2}\;ds\\
&=& \int_0^\tau\left(16C^2\sigma^2+{4 C^2} \;\frac{(1-e^{2{C}\tau})^2}{(1+e^{C\tau})^2}\right)^{1/2}\;ds\\
&=& \int_0^\tau\left(16C^2\;\frac{e^{2{C}\tau}}{(1+e^{C\tau})^2}+{4 C^2} \;\frac{(1-e^{2{C}\tau})^2}{(1+e^{C\tau})^2}\right)^{1/2}\;ds\\
&=& 2C\int_0^\tau\frac{\left(4\;{e^{2{C}\tau}}+(1-e^{2{C}\tau)^2}\right)^{1/2}}{(1+e^{C\tau})}\;ds\\
&=& 2C\tau.
\end{eqnarray*}
Therefore, if we consider the difference of arc-length between 
$\alpha_{C_x+\delta,C_y,r}$ and $\alpha_{C_x,C_y,r}$,   it diverges when $\tau\to\infty$. Therefore, as in the example of the previous section, two nearby geodesics could differ significantly in time.

Another useful indicator of dynamical chaoticity is given by the average volume elements on ${M}_{G}$. The volume element on ${M}_{G}$ is given by
\begin{eqnarray*}
d{ V}_{g}&=& \left(\frac{4}{\sigma^6(1-r^2)}\right)^{1/2}d\mu_x\wedge d\mu_y \wedge d\sigma.
\end{eqnarray*}
Then the volume of an extended region of  ${M}_{ G}$ is 
$$ 
\Delta V_{{M}_{ G}}(\tau)=\int_{\mu_x(0)}^{\mu_x(\tau)}\int_{\mu_y(0)}^{\mu_y(\tau)}\int_{\theta(0)}^{\theta(\tau)}d{ V}_{g}=\frac{-C_xC_y}{4\sqrt{1-r^2}}\frac{\left(1-e^{2C\tau}\right)^4}{e^{2C\tau}(1+e^{2C\tau})^2},$$
and the average volume is
$$
\frac1{\tau}\int_0^{\tau} \Delta V_{{M}_{G}}(t) \;dt = \frac{-C_xC_y}{8C\sqrt{1-r^2}}\left( \frac{e^{2C\tau}- e^{-2C\tau}}{\tau} - \frac{16}{\tau(1 + e^{2C\tau})} - 12 C+\frac{8}{\tau}\right).
$$
Note, that the asymptotic behaviour of the average volume has a regime of diffusive evolution that increase exponentially when $\tau\to\infty$. This behaviour is similar to the one obtained in \cite{peng} (Section 6) for  the model with one Gaussian variable. Also, it is interesting to note that the diffusive behaviour depends on the correlation $r$ (through the constant $C$) but the asymptotic behaviour does not change even when the two particles do not interact between them (i.e., $r=0$).

Finally, we consider the parameter family of neighbouring geodesics 
$$\alpha_{C_x,C_y,r}=\left\{\mu_x(\tau;C_x,C_y,r),\;\;\mu_y(\tau;C_x,C_y,r),\;\;\sigma(\tau;C_x,C_y,r)\right\}$$
where $\mu_x(\tau;C_x,C_y,r),\mu_y(\tau;C_x,C_y,r)$ and  $\sigma(\tau;C_x,C_y,r)$ are given in (\ref{sol}). The Jacobi field equations are:

\begin{eqnarray}\label{jlcg}
\frac{\partial^2J_x}{\partial^2\tau}\!\!\!\!\!\!\!\!\!\!\!&&-\frac{2}{\sigma}\frac{\partial\sigma}{\partial\tau}\frac{\partial J_x}{\partial\tau}\!\!-\!\! 2C_x\sigma\frac{\partial J_\sigma}{\partial\tau}\!\!+\!\!\left(C\sigma^2\!\!+\!\!(C_x\!+\!C_y)\frac{\partial\sigma}{\partial\tau}\right)J_x+\!\!C_x\!\!\left(\frac{rC_y\!-\!C_x}{4(1-r^2)}\sigma^2+\frac{\partial\sigma}{\partial\tau}\right)J_\sigma=0\label{jlcg1}\\
\frac{\partial^2J_y}{\partial^2\tau}\!\!\!\!\!\!\!\!\!\!\!&&-\frac{2}{\sigma}\frac{\partial\sigma}{\partial\tau}\frac{\partial J_y}{\partial\tau}\!\!-\!\! 2 C_y\sigma\frac{\partial J_\sigma}{\partial\tau}\!\!+\!\!\left(C\sigma^2\!\!+\!\!(C_x\!+\!C_y)\frac{\partial\sigma}{\partial\tau}\right)J_y+\!\!C_y\!\!\left(\frac{rC_x\!-\!C_y}{4(1-r^2)}\sigma^2+\frac{\partial\sigma}{\partial\tau}\right)J_\sigma=0\label{jlcg2}\\
\frac{\partial^2J_\sigma}{\partial^2\tau}\!\!\!\!\!\!\!\!\!\!\!&&+\frac{\sigma}{2(1+r)}\left(C_x\!\frac{\partial J_x}{\partial\tau}\!+\!C_y\!\frac{\partial J_y}{\partial\tau}\right)-\frac{2}{\sigma}\frac{\partial\sigma}{\partial\tau}\frac{\partial J_\sigma}{\partial\tau}+\left(\frac{(\frac{J_\sigma}{\partial\tau})^2}{\sigma^2}\!\!+\!\!{(C_x+C_y)}\frac{\partial\sigma}{\partial\tau}\right)J_\sigma\nonumber\\
\!\!\!\!\!\!\!\!\!\!\!&&+\frac{1}{4(1-r^2)}\left(\frac{C_x^2-rC_y^2+(r^2-2)C_xC_y}{4}\sigma^2\!\!+\!\!{(C_x-rC_y)}\frac{\partial\sigma}{\partial\tau}\right)J_x\nonumber\\
\!\!\!\!\!\!\!\!\!\!\!&&+\frac{1}{4(1-r^2)}\left(\frac{C_y^2-rC_x^2+(r^2-2)C_xC_y}{4}\sigma^2\!\!+\!\!{(C_y-rC_x)}\frac{\partial\sigma}{\partial\tau}\right)J_y=0\label{jlcg3}
\end{eqnarray}
 The technical details of this computation can be found in  the Appendix.  
Note that $\displaystyle\lim_{\tau\to\infty}\sigma(\tau)=\displaystyle\lim_{\tau\to\infty}\frac{\partial\sigma(\tau)}{\partial \tau}=0$ and $
\displaystyle\lim_{\tau\to\infty}\frac{\frac{\partial\sigma(\tau)}{\partial\tau}}{\sigma(\tau)}=-C$.
 Therefore, if we assume as in \cite{peng} that
\begin{eqnarray}\label{assump}
\displaystyle\lim_{\tau\to\infty}\sigma(\tau)\frac{\partial J_\sigma}{\partial\tau}=\displaystyle\lim_{\tau\to\infty}\sigma(\tau)\frac{\partial J_x}{\partial\tau}=\displaystyle\lim_{\tau\to\infty}\sigma(\tau)\frac{\partial J_y}{\partial\tau}=0\\
\displaystyle\lim_{\tau\to\infty}\frac{\partial\sigma(\tau)}{\partial\tau}J_\sigma=\displaystyle\lim_{\tau\to\infty}\frac{\partial\sigma(\tau)}{\partial\tau}J_x=\displaystyle\lim_{\tau\to\infty}\frac{\partial\sigma(\tau)}{\partial\tau}J_y=0,\nonumber
\end{eqnarray}
 and we consider the asymptotic limit as $\tau\to \infty$, the Jacobi field equations become,
\begin{eqnarray}\label{jlcg}
\frac{\partial^2 J_x}{\partial^2\tau}\!\!\!\!\!\!\!\!\!\!\!&&+2C\frac{\partial J_x}{\partial\tau}=0,\nonumber \\
\frac{\partial^2 J_y}{\partial^2\tau}\!\!\!\!\!\!\!\!\!\!\!&&+2C\frac{\partial J_y}{\partial\tau}=0,\label{jlcasym}\\
\frac{\partial^2 J_\sigma}{\partial^2\tau}\!\!\!\!\!\!\!\!\!\!\!&&+2C\frac{\partial J_\sigma}{\partial\tau}+C^2J_\sigma=0\nonumber.
\end{eqnarray}
In this case (\ref{jlcasym}) can be easily solved. Thus, the asymptotic solutions are given by
\begin{eqnarray}\label{soluciones}
J_x(\tau)=a_{x,1}+a_{x,2}e^{-2C\tau}\quad\quad J_y(\tau)=a_{y,1}+a_{y,2}e^{-2C\tau}\quad\quad J_\sigma(\tau)=(a_{\sigma,1}+a_{\sigma,2\tau)}e^{-C\tau},
\end{eqnarray}
where $a_{x,j},a_{y,j}$ and $a_{\sigma,j1}$ for $j=1,2$ are real integration constants. Hence, we have that the square norm of the Jacobi field is
\begin{eqnarray*}
\|J\|^2\!\!\!\!&=&\!\!\!\!\frac{1}{\sigma^2(1-r^2)}\!\!\left(\left(a_{x,1}+a_{x,2}e^{-2C\tau}\right)^2\!+\!\left(a_{y,1}+a_{y,2}e^{-2C\tau}\right)^2\!-\!2r\left(a_{x,1}+a_{x,2}e^{-2C\tau}\right)\left(a_{y,1}+a_{y,2}e^{-2C\tau}\right)\right)\\
\!\!\!&+&\!\!\!\!\frac{4}{\sigma^2}(a_{\sigma,1}+a_{\sigma,2\tau})^2e^{-2C\tau}.
\end{eqnarray*}

\noindent\bf Remark \ref{gaussiana}.1: \rm Note that the solution $\sigma(\tau)$ given in (\ref{sol}), satisfies $\lim_{\tau\to\infty}\sigma(\tau)=\lim_{\tau\to\infty}\frac{\partial\sigma(\tau)}{\partial\tau}= 0$. Even more,  $\sigma(\tau)$ and $\frac{\partial\sigma(\tau)}{\partial\tau}$ are asymptotically equivalent to $e^{-2C\tau}$. Therefore, the assumptions given in (\ref{assump}) are satisfied if  $J_x,\ J_y,\ J_{\sigma},\ \frac{\partial J_x}{\partial\tau},\ \frac{\partial J_y}{\partial\tau},$ and $  \frac{\partial J_\sigma}{\partial\tau}$ are  of order $o(e^{2C\tau})$. For instance, notice that the solutions obtained in (\ref{soluciones}) satisfy these assumptions.

Finally, it is easy to see that the main term of the asymptotic expansion has  exponential behaviour equivalent to $2\left(\frac{a_{1,x}^2+a_{1,y}^2-2 r a_{1x}a_{1,y}}{1-r^2}\right)e^{2C\tau} $. Note that when the particles do not interact between them, the Jacobi field has the same asymptotic behaviour. On the other hand, if the particles are strongly   related (i.e., $r\to 1$) the norm of the Jacobi field goes to infinity.

\section{Final remarks and conclusions}

In this paper, we investigate two entropic dynamical models corresponding to statistical manifolds with different characteristics.
The first one corresponds to a system of four uncorrelated particles in a one dimensional space modelled by a statistical manifold of $4-$joint one parameter exponential density. The second one describes the behaviour of two particle interacting between them according to a multivariate Gaussian distribution. For both models, we study their geometric structure from the viewpoint of information geometry. In order to analyse the character of the  stability for both models, we obtain explicit parametrizations of  the geodesics and we study their behaviour. Also, we compute the volume of an extended region of each manifold and the Jacobi field associated with the geodesic deviation equations on the manifolds. We concluded that both models show clear signs of instability.

Finally, we want to note that if we combine the studied models, we  obtain a large class of statistical manifolds that can be analysed easily using  the results obtained here. More precisely, assume that we have the following statistical manifold
$${M}=\left\{  p_1((x_1,x_2,x_3,x_4),(\theta_1,\theta_2,\theta_3,\theta_4))\;p_2((x_5,x_6),(\mu_{x_6},\mu_{x_7},\sigma)) \quad \theta_i>0 \mbox{ for } i=1,\dots,4 \mbox{ and } \sigma>0 \right\}$$
where $p_1$ and $p_2$ are defined as in (\ref{modelo1}) and (\ref{modelo2}), respectively. $M$ modelled a system of  six particle $\bx=(x_1,x_2,x_3,x_4,x_5,x_6)$ in a one dimensional space with no interaction   between them,  except $(x_5,x_6)$. We will not stop here in the details, however it is easy to see that  the scalar curvature of ${M}$ is $\frac{-3}{2}$, the geodesic equations correspond to a system of seven equations given by (\ref{geodesicmodel1}) and (\ref{eqg}) and  the square norm of the Jacobi field  is $\|J\|^2=\|J_{{ M}_4}\|^2+\|J_{{M}_{G}}\|^2$,  where $J_{{ M}_4}$ and $J_{{M}_{G}}$ are the Jacobi fields of $M_4$ and $M_G$, respectively. The negative sign of the scalar curvature and the exponential grow of  $\|J\|^2$ show local instability of this system.

The progress presented in this work constitute an advance for characterize the chaos of the entropic dynamical models and extend the important results obtained in \cite{peng}.

\section*{Appendix}

In this section we sketch  the steps to get  the equations (\ref{jlcg}). First step, we compute $R^i_{klm}$
\begin{eqnarray*}
R^1_{111}&=&R^1_{122}=R^2_{222}=R^2_{211}=\frac{1}{4\sigma^2(1-r^2)}=-R^1_{131}=-R^2_{232},\\
R^1_{121}&=&R^1_{112}=R^2_{212}=R^2_{221}=\frac{-r}{4\sigma^2(1-r^2)}=-R^1_{132}=-R^2_{231},\\
R^1_{311}&=&R^1_{133}=R^1_{312}=R^2_{322}=R^2_{233}=R^2_{321}=R^3_{331}=R^3_{332}=\frac{1}{\sigma^2}=-R^1_{331}=-R^1_{313}=-R^2_{332}=-R^2_{323},\\
R^3_{121}&=&R^3_{112}=R^3_{212}=R^3_{221}=\frac{-r}{16\sigma^2(1-r^2)},\\
R^3_{111}&=&R^3_{222}=\frac{1}{16\sigma^2(1-r^2)},\\%\quad\quad\quad\quad\quad\quad\quad\quad R^3_{331}=R^3_{332}=\frac{-r}{2\sigma^2(1-r^2)}\\
R^3_{211}&=&R^3_{122}=\frac{r^2}{16\sigma^2(1-r^2)},\\%\quad\quad\quad\quad\quad\quad\quad\quad R^3_{321}=R^3_{312}=\frac{-r}{2\sigma^2(1-r^2)}.
R^3_{331}&=&R^3_{332}=\frac{-r}{2\sigma^2(1-r^2)},\\
R^3_{321}&=&R^3_{312}=\frac{-r}{2\sigma^2(1-r^2)}.
\end{eqnarray*}
Therefore using the fact that $\frac{\partial \mu_x(\tau)}{\partial\tau}=C_x\sigma^2$ and $\frac{\partial \mu_y(\tau)}{\partial\tau}=C_y\sigma^2$, the Jacobi field equations reduce to
\begin{eqnarray}\label{eqjlc}
\frac{D^2J_x}{D\tau^2}&+&J_x\left(\sigma^2\frac{C_x^2-rC_xC_y}{4(1-r^2)}+\frac{\partial \sigma}{\partial\tau}(C_x+C_y)-\left(\frac{\frac{\partial \sigma}{\partial\tau}}{\sigma}\right)^2\right)\nonumber\\
&+&J_y\left(\sigma^2\frac{C_xC_y-rC^2_x}{4(1-r^2)}\right)+J_\sigma\left(\sigma^2\frac{-C_x^2+rC_xC_y}{4(1-r^2)}\right)=0,\nonumber\\
\frac{D^2J_y}{D\tau^2}&+&J_y\left(\sigma^2\frac{C_y^2-rC_xC_y}{4(1-r^2)}+\frac{\partial \sigma}{\partial\tau}(C_x+C_y)-\left(\frac{\frac{\partial \sigma}{\partial\tau}}{\sigma}\right)^2\right)\\
&+&J_x\left(\sigma^2\frac{C_xC_y-rC^2_y}{4(1-r^2)}\right)+J_\sigma\left(\sigma^2\frac{-C_y^2+rC_xC_y}{4(1-r^2)}\right)=0,\nonumber\\
\frac{D^2 J_\sigma}{D\tau^2}&+&J_x\left(\sigma^2\frac{C_x^2+(r^2-r)C_xC_y-rC_y^2}{16(1-r^2)^2}+\frac{\partial \sigma}{\partial\tau}\frac{C_x-rC_y}{2(1-r^2)}\right)\nonumber\\
&+&J_y\left(\sigma^2\frac{C_y^2+(r^2-r)C_xC_y-rC_x^2}{16(1-r^2)^2}+\frac{\partial \sigma}{\partial\tau}\frac{C_y-rC_x}{2(1-r^2)}\right)\nonumber\\
&+&J_\sigma\frac{\partial \sigma}{\partial\tau}(C_x+C_y)=0.\nonumber
\end{eqnarray}
The expression of the second derivative of $J_x, J_y$ and $J_\sigma$ are 
\begin{eqnarray}\label{eqdc}
\frac{D^2J_x}{D\tau^2}=&-&2 \frac{\frac{\partial \sigma}{\partial\tau}}{\sigma}\frac{\partial J_x}{\partial\tau}-2C_x\sigma\frac{\partial J_\sigma}{\partial\tau}+J_x\left(-\frac{\frac{\partial^2 \sigma}{\partial\tau^2}}{\sigma}+2\left(\frac{\frac{\partial \sigma}{\partial\tau}}{\sigma}\right)^2-\sigma^2\frac{C_x^2-rC_xC_y}{4(1-r^2)}\right)\nonumber\\
&+&J_y\left(\sigma^2\frac{rC_x^2-C_xC_y}{4(1-r^2)}\right)+J_\sigma\frac{\partial \sigma}{\partial\tau}Cx,\nonumber\\
\frac{D^2 J_y}{D\tau^2}=&-&2 \frac{\frac{\partial \sigma}{\partial\tau}}{\sigma}\frac{\partial J_y}{\partial\tau}-2C_y\sigma\frac{\partial J_\sigma}{\partial\tau}+J_y\left(-\frac{\frac{\partial^2 \sigma}{\partial\tau^2}}{\sigma}+2\left(\frac{\frac{\partial \sigma}{\partial\tau}}{\sigma}\right)^2-\sigma^2\frac{C_y^2-rC_xC_y}{4(1-r^2)}\right)\\
&+&J_x\left(\sigma^2\frac{rC_y^2-C_xC_y}{4(1-r^2)}\right)+J_\sigma\frac{\partial \sigma}{\partial\tau}Cy,\nonumber\\
\frac{D^2J_\sigma}{D\tau^2}=&-&2 \frac{\frac{\partial \sigma}{\partial\tau}}{\sigma}\frac{\partial J_\sigma}{\partial\tau}
+\frac{\sigma(1-r)C_x}{2(1-r^2)}\frac{\partial J_x}{\partial\tau}+\frac{\sigma(1-r)C_y}{2(1-r^2)}\frac{\partial J_y}{\partial\tau}\nonumber\\
&+&J_\sigma\left(\frac{\frac{\partial \sigma}{\partial\tau}}{\sigma}\right)^2+\frac{1}{4(1-r^2)} \frac{\partial \sigma}{\partial\tau}J_x+\frac{1}{4(1-r^2)}\frac{\partial \sigma}{\partial\tau}J_y.\nonumber
\end{eqnarray}
Finally the equations (\ref{jlc}) follow from (\ref{eqjlc}) and (\ref{eqdc}).

\end{document}